# A Magnon Scattering Platform

**Authors:** Tony X. Zhou[1,2,†], Joris J. Carmiggelt[1,3,†], Lisa M. Gächter[1,4,†], Ilya Esterlis[1], Dries Sels[1], Rainer J. Stöhr[1,5], Chunhui Du[1,6,8], Daniel Fernandez[1], Joaquin F. Rodriguez-Nieva[1], Felix Büttner[7], Eugene Demler[1] and Amir Yacoby[1,2,*].

**Affiliations**
[1] Department of Physics, Harvard University, 17 Oxford Street, Cambridge, Massachusetts 02138, USA.

[2] John A. Paulson School of Engineering and Applied Sciences, Harvard University, Cambridge, Massachusetts 02138, USA.

[3] Department of Quantum Nanoscience, Kavli Institute of Nanoscience, Delft University of Technology, Lorentzweg 1, 2628 CJ Delft, The Netherlands.

[4] Solid State Physics Laboratory, ETH Zurich, 8093 Zurich, Switzerland.

[5] Center for Applied Quantum Technology and 3rd Institute of Physics, University of Stuttgart, 70569 Stuttgart, Germany.

[7] Department of Materials Science and Engineering, Massachusetts Institute of Technology, Cambridge, MA 02139, USA.

[8] Department of Physics, University of California, San Diego, La Jolla, California 92093.

* Correspondence to: yacoby@physics.harvard.edu
[†] These authors contributed equally to this work.

**Abstract:** Scattering experiments have revolutionized our understanding of nature. Examples include the discovery of the nucleus, crystallography, and the discovery of the double helix structure of DNA. Scattering techniques differ by the type of the particles used, the interaction these particles have with target materials and the range of wavelengths used. Here, we demonstrate a new 2-dimensional table-top scattering platform for exploring magnetic properties of materials on mesoscopic length scales. Long lived, coherent magnonic excitations are generated in a thin film of YIG and scattered off a magnetic target deposited on its surface. The scattered waves are then recorded using a scanning NV center magnetometer that allows sub-wavelength imaging and operation under conditions ranging from cryogenic to ambient environment. While most scattering platforms measure only the intensity of the scattered waves, our imaging method allows for spatial determination of both amplitude and phase of the scattered waves thereby allowing for a systematic reconstruction of the target scattering potential. Our experimental results are consistent with theoretical predictions for such a geometry and reveal several unusual features of the magnetic response of the target, including suppression near the target edges and gradient in the direction perpendicular to the direction of surface wave propagation. Our results establish magnon scattering experiments as a new platform for studying correlated many-body systems.

**Main Text:**

Scattering experiments use a coherent source of waves or particles that impinge on a specimen with well-defined energy and momentum. The scattered waves form a unique fingerprint of the specimen that can then be used to reconstruct certain underlying material properties. For example, optical scattering has provided deep insight into the underlying dielectric response of materials and has enabled the exploration of dipole-coupled excitations such as excitons (*1*), polaritons (*2*), and nonlinear optical phenomena (*3*). At short wavelengths, x-ray scattering can reveal the underlying atomic structure of materials, and neutron scattering provides the ability to study magnetic order down to the atomic scale (*4*). Often, however, existing scattering methods require large quantities of material in order to have an appreciable scattering intensity. Materials such as 2-dimensional (2D) layered materials, for example, pose a severe challenge to traditional scattering platforms since they are only a few monolayers thick and typically only a few micrometers wide. Developing alternative table-top scattering techniques for mesoscopic samples is therefore required. Here, we demonstrate for the first time a table-top scattering platform (Fig.

1A) that uses coherent magnonic waves as the impinging particles. To establish this as a new scattering platform we need to demonstrate: 1) The ability to launch coherent waves with well defined energy and momentum; 2) Accurate detection of scattered waves, ideally, both amplitude and phase; 3) Show that we have appreciable interaction of magnons with the target material; and 4) Achieve reliable extraction of target material properties. Below, we demonstrate that we have successfully achieved all of these goals.

*Launching Magnons* - Since magnons cannot propagate in free space, we use a thin film of Yttrium Iron Garnet (YIG) as the 'vacuum' supporting coherent long lived magnonic excitations with well-defined energy and momentum. Coherent generation of magnons in YIG is well established scientifically (*5*) with high degree of control and tunability in phase, amplitude, and wavelength spanning several nanometers to hundreds of micrometers. We generate magnons in YIG using a micro stripline deposited on the surface of a 100 nm thick YIG grown on $Gd_3Ga_5O_{12}$ (GGG) substrate (*6*) (Fig. 1A). By driving a microwave current through the stripline, coherent magnonic excitations are launched at the frequency of the microwaves and at a wavelength set by the underlying magnonic dispersion, $\omega(k)$, in YIG (*7*). In the presence of an external magnetic field pointing along the stripline (Fig. 1B, left), this geometry launches magnons with k-vector perpendicular to the microwave current direction.

*Detection of scattered waves* - A key component of a scattering platform is the ability to image the scattered waves. Currently, there are several established techniques for imaging magnons in YIG including Brillouin light scattering (*8*, *9*), optical Kerr microscopy (*10*), and resonant X-ray microscopy (*11*). Here, we demonstrate the use of a single nitrogen vacancy (NV) center in diamond as a local sensor for magnonic excitations sensitive to both amplitude and phase with nanometer resolution.

The low-energy manifold of an NV center consists of an S=1 spin triplet. Its ground state corresponds to $m_s = 0$ and its excited states consist of $m_s = \pm 1$ states. At zero magnetic field, the $m_s = 0$ state is split from the excited state by 2.87 GHz. Application of a finite magnetic field along the NV axis splits the $m_s = \pm 1$ states by $2\gamma_e B_{ext}$ allowing for static magnetic field detection (Fig. 1B, right). As a scanning probe, NV center microscopy (*12*, *13*) has recently been used to image spin textures of skyrmions (*14*), non-collinear antiferromagnets (*15*), magnetic domains in 2D material (*16*), and viscous current flow (*17*).

Propagating magnons and the scattered magnons due to the target generate local time varying magnetic fields above the YIG. These can be detected by an NV center if the frequency of the magnons matches that of the electron spin resonance (ESR) of the NV center. We use an external magnetic field to tune the ESR frequency of the NV center to match that of the excited magnons (Fig. 1B). Under these conditions, the NV center undergoes transitions from its ground state to one of its excited states (typically the $m_s = -1$ state) and its corresponding fluorescence will reflect the occupation of the NV center. In the $m_s = 0$ state, the fluorescence is strong, and it is weaker in the excited states. Under weak continuous drive, the fluorescence is proportional to the intensity of the driving field which in turn is directly proportional to the amplitude of the magnonic excitation at that location. Fig. 1E shows the fluorescence of an NV center as a function of the magnon frequency and $B_{ext}$. A clear decrease in NV fluorescence can be seen when the magnon frequency matches the NV ESR frequency. When the driving AC magnetic field generated by the magnons is strong and coherent, Rabi oscillation of the NV center can be detected (Fig. 1E, inset). When the excitation frequency matches the ferromagnetic resonance (FMR) of YIG, additional suppression of fluorescence can be observed in Fig. 1E. This effect results from FMR generated magnons and associated magnetic field noise at the NV ESR frequencies (*18*).

We determine the phase of the propagating coherent magnons using an interference scheme (*6*). In the rotating frame of the NV center, the phase of the oscillating field determines the axis along which the spin rotates (Fig. 1C). To determine this axis, and hence the phase, we apply another RF reference field that is uniform in space, both in amplitude and phase, and has the same ESR frequency. This is achieved using a wire antenna situated several tens of micrometers away from the sample (loop in Fig. 1A). The total AC field driving the NV center is a vector sum of the field generated locally by the magnon and the reference field:

$$B_{total} = B_{magnon} + B_{ref} = Re\{e^{i(kx-\omega t)} + e^{i(-\omega t+\varphi)}\} = Re\{[e^{ikx} + e^{i\varphi}](e^{-i\omega t})\} \quad (1)$$

Here, k is the wavenumber, ω is the drive frequency and φ is the phase difference between reference field and magnon field. The amplitude of both signals is normalized to 1 for intuitive illustration (Fig. 1, C and D). By scanning the NV probe across the sample, we observe an increase of fluorescence at locations where the magnon field is exactly out of phase with the reference field (Fig. 2A). These peaks in fluorescence recur each time we move a distance corresponding to one wavelength of the magnons. As we vary the phase difference of the two microwave sources we are able to capture the real space propagating component of the magnons (Fig. 2B). A full movie can be seen in supplementary information (*6*). Extracting the wavelength of magnons as a function of frequency allows us to directly extract the dispersion relation of the magnons (Fig. 2, C and F). We determine the dispersion down to a wavelength of 640 nm (Fig. 2D and E) limited only by the inefficient generation of magnons by the stripline at shorter wavelengths described in detail below. Even with this simple RF waveguide design (*6*), we nevertheless are on par with the shortest magnon wavelength detected using visible optical techniques (*19–22*).

We first employ our NV magnon detection scheme to characterize the generation of magnons from the stripline (23, 24). At a given drive frequency, ω, only magnons of wavevector k that satisfy the dispersion relation of the magnetic medium are launched. However, the excitation efficiency associated with a particular k is also set by the spatial geometry of the stripline. While magnons with wavelengths larger than the width of the stripline can be easily excited, this is not true for magnons with wavelengths considerably shorter than this width. The excitation efficiency of the stripline as a function of k is governed by the normalized Fourier transform of the spatial microwave field generated by the rectangular stripline, $H(k)$ (Fig. 3A). Besides the fact that NV center measures the magnetic field generated by magnons of this wavevector, however, our measured value is further scaled by $D(k,z) \propto ke^{-kz}$ (25) also known as the filter function. Both $H(k)$ and $D(k,z)$ are shown in Fig. 3A. The total measured magnetic field is therefore expected to be given by:

$$|B_{magnon}| = A \times D(k,z) \times H(k) \qquad (2)$$

where $A$ is a pre-factor accounting for the NV axis oriented at an angle relative to the YIG and z is the distance between NV and YIG surface (Fig 1B, left inset). Fig. 3D shows both ESR and Rabi measurements due to magnons excited at different frequencies. The oscillatory function due to $H(k)$ is clearly visible. Fig. 3B shows the ESR response along the NV ESR line taken at different NV heights above the YIG. A shift in weight of the ESR spectrum to lower k is visible for higher z in accordance with the expected NV filter function. Ultimately, what we are doing in the section is to experimentally determine the point spread function of the "detector" to be used in following scattering experiment.

*Interaction of magnons with a target material* - We now turn to describe the interaction of magnonic waves with a target material. For our target, we use a 100-nm thick Py disk with 5-micron diameter deposited directly onto the YIG. Coherent magnonic plane waves are launched using our microwave stripline. Upon impinging on the Py disk as described below, magnons are scattered, and the coherent sum of the scattered and unscattered magnons is measured (Fig. 4A). The modulation in intensity observed suggests that the scattering of magnons is a coherent, inelastic process. Additionally, we find that the scattered magnons are confined within an angular opening $2\theta_c$ which is a direct consequence of the chiral nature of the magnons excited (*26–28*). A second scattering map containing information on the local scattering phase is generated by applying an additional RF signal from the distant antenna (Fig. 4D). The RF field due to the antenna interferes with the magnetic field generated by the scattered and unscattered waves thereby producing an image that encodes information about the local phase of the scattered waves.

*Reliable extraction of target material properties* - The most prominent features in the wave pattern of Fig. 4 A and D are (i) negligible back scattering and (ii) confinement of the scattered wave to a cone ahead of the target. These qualitative features of the scattered wave are determined entirely by the dispersion relation of the Damon-Eshbach surface waves (DESW), as has been extensively studied in (*5, 29–32*). The negligible backscattering is due to the "field displacement nonreciprocity" of the free DESW, which implies that waves will be localized on either the top or bottom surfaces of the magnetic film, depending on the direction of propagation. The scattering cone is a direct consequence of the specific dispersion relation of the DESW. The isofrequency curves of the DESW dispersion asymptote to a cone in momentum space whose opening angle is given by:

$$\theta_c = \sin^{-1}\left(\frac{\omega+\sqrt{\omega^2-\omega_0(\omega_0+\omega_M)}}{\omega_0+\omega_M}\right). \tag{3}$$

Here, $\omega$ is the mode frequency, $\omega_0 = \gamma_e B_{ext}$ is and $\omega_M = \gamma_e M_S$ (*6*). The group velocity $\mathbf{v}_g$ is normal to isofrequency curves, and energy flow will therefore be limited to a cone in real space, with opening angle $\theta_c$ with respect to the $x$-axis. For the frequency shown in Fig. 4A ($f = 2.18$ GHz), the opening angle obtained using Eq.3 is $\theta_c \approx 34°$. This is in reasonable agreement with the measured opening angle, $\theta_c \approx 28° \pm 2°$. Furthermore, it can be shown (*29*) that the mode density of DESW diverges upon approaching $\theta_c$, which explains the precipitous rise in the amplitude of the scattered wave observed near the critical angle in Fig. 4 A and D. Beyond these gross features, an intricate structure in the phase profile and contrast of the scattered wave can be easily discerned. These details provide unique and detailed information about the target which we exploit below.

To provide a quantitative model of the magnon scattering by the Py disk, we add a spatially bounded AC source term to the wave equation for the DESW. We note that this simple model ignores any DC coupling between the target and the YIG. Physically, this means that we ignore modifications of the local magnetic permeability seen by the magnons in the vicinity of the target – i.e., we neglect effects of the target on the magnon "vacuum." We then compute the Green's function for this wave equation, from which we can determine the scattered wave for an arbitrary source. Conversely, equipped with the Green's function and experimental wave patterns, we can invert the problem to determine the scattering potential. In the spirit of conventional scattering experiments, we adopt the latter approach to interrogate magnetic properties of the target.

In general, the scattered wave is given by the convolution $B_{s\alpha}(\mathbf{r}) = \int_{r'} d^3\mathbf{r}' \, G(\mathbf{r}-\mathbf{r}')\,\partial_\alpha[\nabla \cdot \mathbf{m}(r')]$, where $G(\mathbf{r})$ is the Green's function, $\nabla \cdot \mathbf{m}$ is the source, and subscript $\alpha$ denotes a direction orthogonal to the NV axis. In principle, given $B_{s\alpha}$ and $G$, the source term is completely determined by inverting the convolution. However, for the current experimental setup we find that the presence of significant background as well as noise prohibit carrying out this procedure explicitly (*6*). The inversion is therefore done by fitting the "source", i.e. $\nabla \cdot \mathbf{m}$, to best match the intensity pattern seen in the experiment. As basis functions for the fit we take the Gaussian function $e^{-(x^2+y^2)/\sigma^2}$ and its derivatives up to 8th order in both directions, where σ is fixed to be the radius of the Py disk. We separate the source into real and imaginary components (Fig. 4 B and C, left and right insets), to reflect a possible phase shift of the scattered wave with respect to the incident wave. Results of the analysis for the phase-resolved images are presented in Figure 4. Note that we shifted the overall phase of the source in a way that makes it easier to separate components with different symmetries. The simplest model of magnetization dynamics of the target excited by the DESW corresponds to magnetic moment of the disk precessing around the direction of the static field, i.e. in the xz-plane. Since the disk is very thin in the z-direction, the dominant contribution to $\nabla \cdot \mathbf{m}$ should come from the x-derivative of the x-component of magnetization. This should produce a dipolar pattern for the source aligned along the x-axis. Figure 4B, right inset shows that the imaginary part of the source is indeed of the dipolar type, and it is almost an order of magnitude larger than the real part (Fig. 4 B, left inset). The real part of the source has a quadrupolar character which implies that the source has an additional gradient along the y-axis. We remind the readers that incident wave is propagating along the x-axis, hence this y-gradient cannot be related to the spatial profile of the incident wave. It appears to be an intrinsic feature of the target revealed by our scattering experiment. We comment on several other interesting features of our analysis. While we achieve excellent fit of our model to the phase images, using the same parameters for the corresponding amplitude images does not automatically produce a good fit. We expect that understanding this discrepancy requires introducing a more sophisticated version of the Py disk/YIG interaction, including renormalization of static susceptibility parameters of YIG below the target. Secondly, we find the best agreement between theoretically simulated phase images and the experimental data when the target is represented using Gaussian based functions, rather than functions with sharper edges. This suggests that response of the Py target near the disk edges is strongly suppressed. Finally, experiments have been done in the regime where the diameter of the target is of the order of the magnon wavelength. One can then expect the disc to develop spatial gradients in the direction of propagation of DESW. Surprisingly, we find that the main response of the target corresponds to magnetization of the entire disk oscillating as a whole.

Our magnonic scattering platform provides a new way for exploring mesoscopic materials. While clearly suited for magnetic target materials, the time varying magnetic fields generated by the magnons can also lead to strong interaction with other phases of matter such as superconductors, topological insulators with conducting surface states, spin liquids and more, thereby providing new insights into such phases that are unattainable by other methods.

## Acknowledgments

This work was primarily supported by the U.S. Department of Energy, Basic Energy Sciences Office, Division of Materials Sciences and Engineering under award DE-SC0001819. A. Y. is also partly supported by ARO Grants No. W911NF-17-1-0023 and the Gordon and Betty Moore Foundation's EPiQS Initiative through Grant No. GBMF4531. Fabrication of samples was supported by the U.S. Department of Energy, Basic Energy Sciences Office, Division of Materials Sciences and Engineering under award DE-SC0019300. A.Y. also acknowledges support from ARO grants W911NF-18-1-0316, and W911NF-1-81-0206. J.C. was supported by the Netherlands Organisation for Scientific Research (NWO/OCW), as part of the Frontiers of Nanoscience program. L.G. was supported by the Zeno-Karl-Schindler Master Thesis Grant. I.E., D.S., J.R-N. and E.D. acknowledge support from Harvard-MIT CUA, AFOSR-MURI Photonic Quantum

Matter (award FA95501610323), DARPA DRINQS program (award D18AC00014), and Harvard Quantum Initiative. D.S. acknowledges support from the FWO as post-doctoral fellow of the Research Foundation Flanders. D.F acknowledges the support by the National Science Foundation under Grant No. EFMA-1542807. Sample fabrication was performed at the Center for Nanoscale Systems (CNS), a member of the National Nanotechnology Coordinated Infrastructure (NNCI), which is supported by the National Science Foundation un-der NSF award no. ECCS -1541959. CNS is part of Harvard University. We thank Mathew Markham and Element Six (UK) for providing diamond samples. We also thank Ronald Walsworth and Matthew Turner for annealing diamonds, and Pablo Andrich and Sungkun Hong for fruitful discussions.

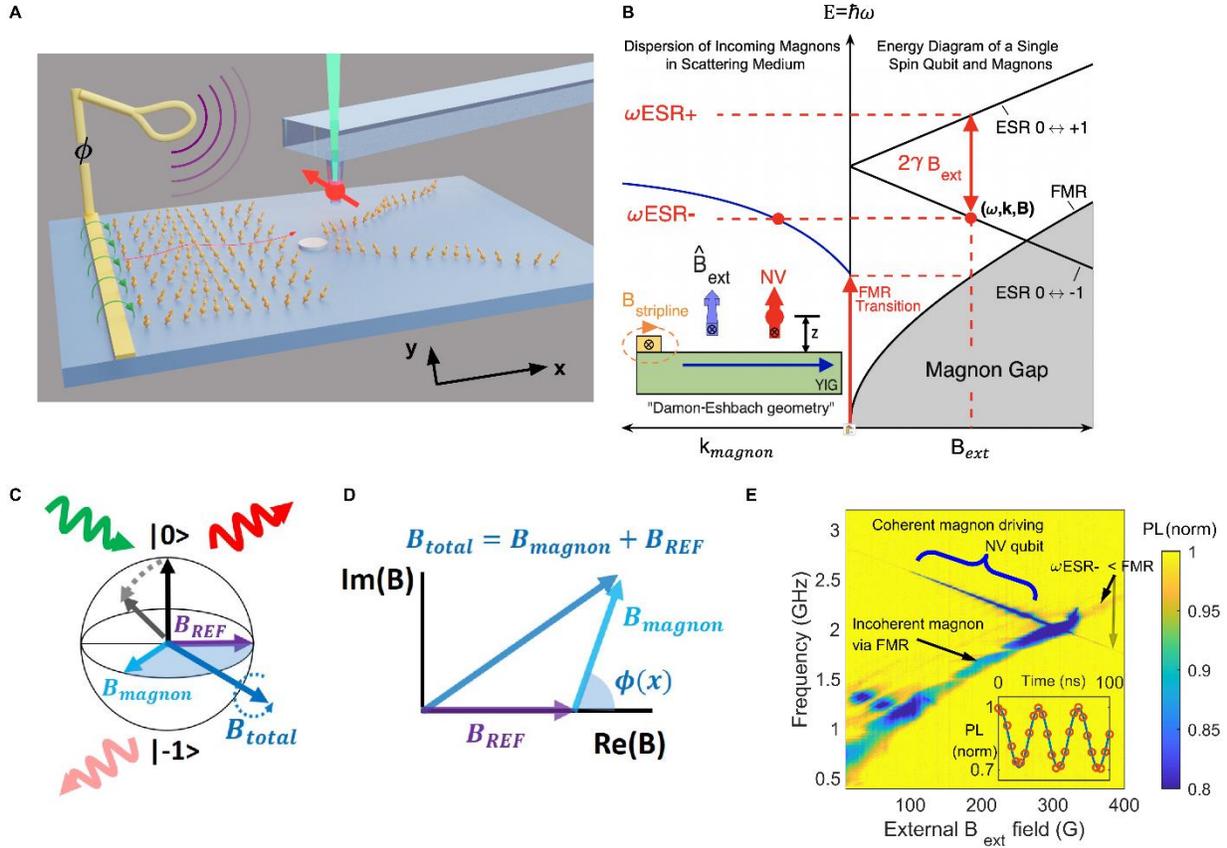

**Fig. 1. Magnon scattering platform and coherent sensing with a single spin magnetometer**

(**A**) Sketch of the magnon-based scattering platform, comprising a microwave stripline as a source, a single NV on a scanned tip as a detector, 100-nm thick YIG as the 'vacuum' supporting long-lived propagating magnons, and a disk-shaped target. The single NV magnetometer allows detection of both amplitude and phase of the scattered magnons. (**B**) Left - Sketch of the magnon dispersion shown in solid blue. Right - NV center energy diagram as function of external field (top) and magnon spin gap (bottom shaded region). For any given magnetic field, there is a unique frequency that matches the NV ESR frequency (e.g. $m_s = 0 \leftrightarrow -1$) and a corresponding magnon with the same frequency and a unique wavevector k determined from the dispersion of YIG, ω(k). Inset: cross-sectional sketch of the YIG, the stripline indicating direction of magnon propagation (blue arrow pointing towards right). The external magnetic field is applied along the NV axis which is oriented parallel to the microwave stripline and tilted upwards along the diamond <111> direction 35.26 degrees out of plane (light blue arrow pointing into the page and slightly upward). **(C)** Bloch sphere representation of the NV spin state under the influence of an AC magnetic field generated by the magnons and reference microwave radiation. The z axis of the sphere is in the direction of NV axis, <111> in diamond. The green arrow represents the green light exciting the NV center and the red arrows represent the different intensity of the emitted red light in each of the spin states of the NV center. **(D)** Schematic phasor representation of the AC magnetic field generated by magnons and reference microwave radiation. **(E)** Normalized fluorescence of the NV center as a function of $B_{ext}$ and frequency. Diminished fluorescence is observed when the excitation matches the ESR frequency of the NV center and along the ferromagnetic resonance. Inset: Observed Rabi oscillation along the NV ESR transition confirming the coherent nature of the field generated by magnon.

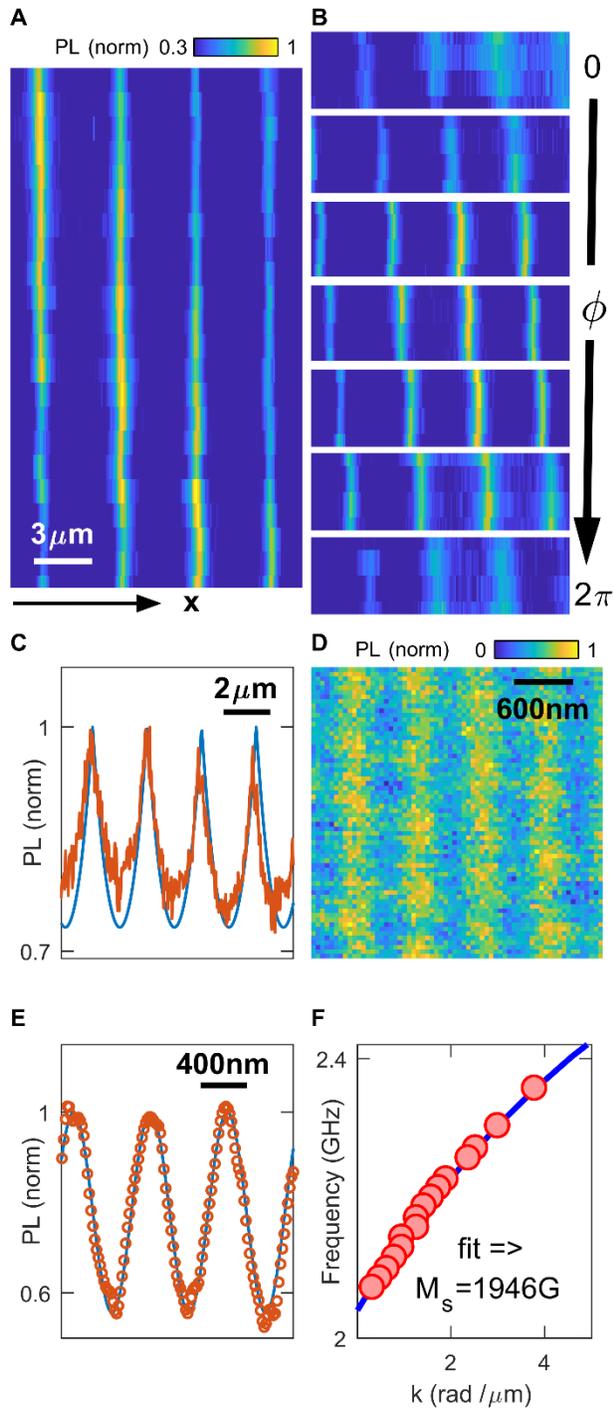

**Fig. 2. Phase imaging of coherent magnons and their dispersion.**

(A) Spatial image of NV fluorescence under continuous drive of both the stripline and remote antenna. The bright fluorescence signal corresponds to destructive interference of the reference RF signal from the antenna and magnon signals (6). (B) Evolution of the magnon wavefront observed by shifting the relative phase (0 to 2π) of the reference source relative to the signal supplied to the stripline. **(C)** Linecut of an interference pattern generated with magnon frequency at 2.3 GHz corresponding to a wavelength of 2.35 µm. **(D)** Imaging magnons with short wavelength. Magnons with wavelength down to 660 nm can easily be resolved. **(E)** Line average of image from (D). **(F)** Magnon dispersion extracted from the fluorescence phase maps.

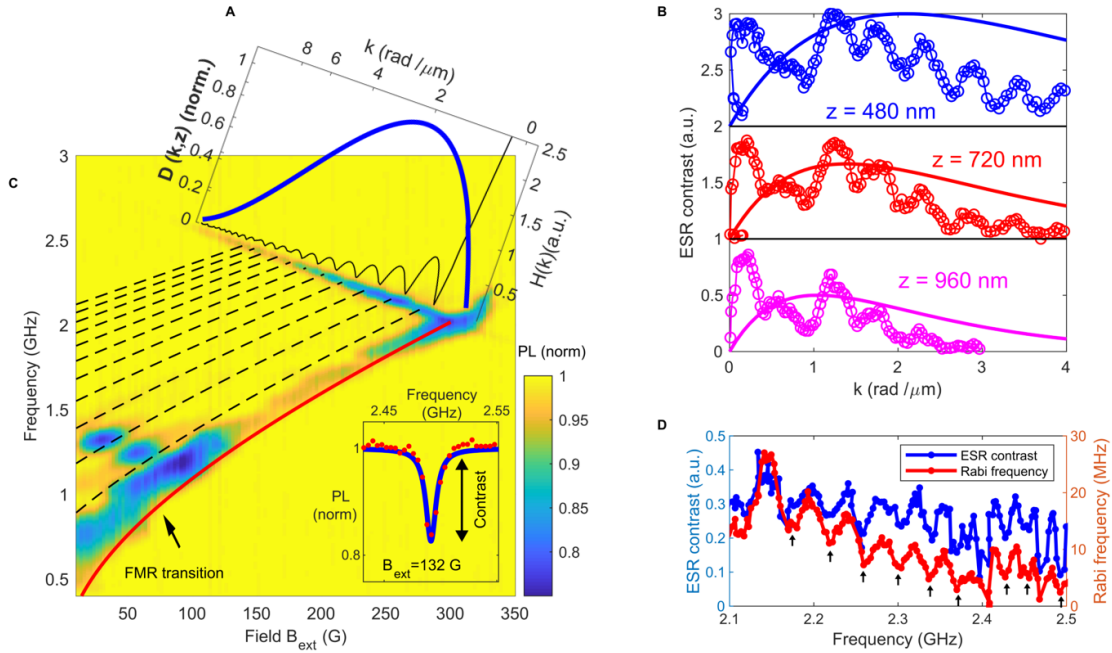

**Fig. 3. Characterization of magnons generated by the microwave stripline**

(**A**) Sketch in k space of the NV filter function $D(k,z)$ (blue, left axis) and magnetic field generated by the microwave stripline $H(k)$ (black, right axis). Each k value is uniquely matched by the external magnetic field and the corresponding ESR frequency. (**B**) ESR fluorescence of the NV center as a function of k for various distances z of the NV center above the YIG. A clear oscillation is observed in accordance with the expected behavior of $H(k)$. Solid lines are the predicted filter function $D(k,z)$ (6). (**C**) Normalized fluorescence of the NV center as a function of $B_{ext}$ and frequency. The dashed straight lines correspond to the nodes in the Fourier spectrum of the simulated H(k) (6). Inset: Vertical linecut in the color map showing an ESR measurement at $B_{ext} = 132$ G. Its contrast is directly proportional to magnon field amplitude. (**D**) Detailed ESR and Rabi measurements along the NV ESR transition. Peaks in the oscillations correspond to magnon modes that are excited efficiently by the microwave stripline. Arrows indicate magnonic modes that are inefficiently excited according to our numerical simulation (6).

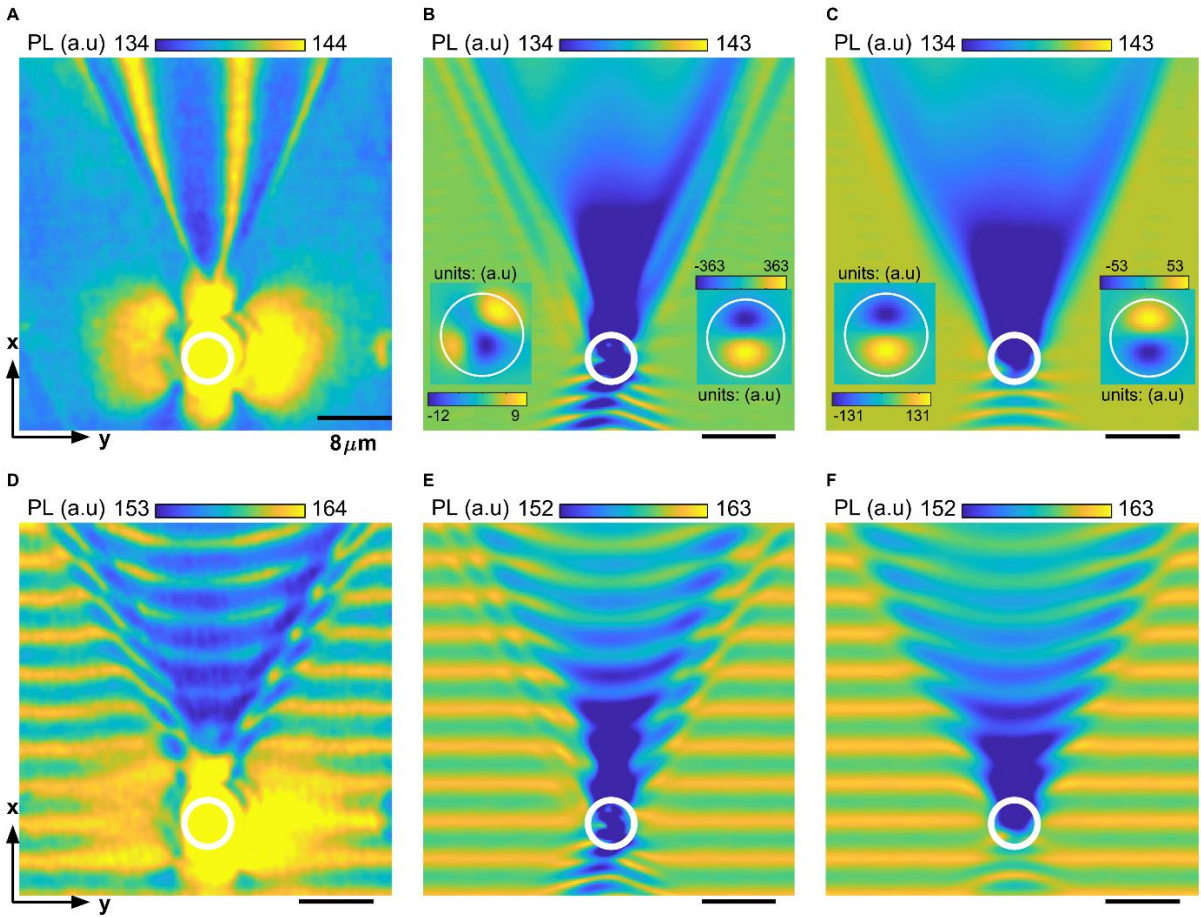

**Fig. 4. Magnon scattering off a target**

Magnons are launched from a microwave stripline on the bottom. While they propagate in the x-direction, they impinge on a Py disk that was deposited on the surface of the YIG (indicated by the white circle). **(A)** The incoming plane wave scatters from the defect and the magnetic field fluctuations caused by the interference of this scattered wave with the incident wave is picked up by the NV. The data is averaged of 39 runs and smoothened over a 100 nm Gaussian window to reduce the noise. Close to the Py disk we observe a "flower" shaped magnetization profile, consistent with static field from a saturated magnetic disk shifting the ESR frequency of NV center to modulate fluoresce. A clear cone is observed, as expected from DESW theory. **(B)** Best fit for a truncated basis set of localized sources. Inset, source image, left: real component, right:

imaginary component. **(C)** Theoretical prediction of the observed intensity if the source would be described by a simple dipole (see inset, left: real component, right: imaginary component). Theoretical model parameters are fit to the data as described in the supplement. **(D)** An additional homogenous microwave field is superimposed on the magnon field. The resulting fringes clearly indicate the plane wave nature of the magnons outside the Damon-Eshbach cone. Additional fringes in the cone provide valuable information about the nature of the scatter. **(E)** Best fit for a truncated set of localized sources. **(F)** Theoretical prediction of the observed intensity in panel D for an optimized dipolar source.